\documentclass{article}

\usepackage{arxiv}

\usepackage[utf8]{inputenc} % allow utf-8 input
\usepackage[T1]{fontenc}    % use 8-bit T1 fonts
\usepackage{hyperref}       % hyperlinks
\usepackage{url}            % simple URL typesetting
\usepackage{booktabs}       % professional-quality tables
\usepackage{amsmath}       % blackboard math symbols
\usepackage{nicefrac}       % compact symbols for 1/2, etc.
\usepackage{microtype}      % microtypography
\usepackage{lipsum}
\usepackage{multirow,bm}
\usepackage{graphicx}
\usepackage{geometry}
\usepackage{array}
\usepackage{algorithm,algorithmic}
\usepackage{xcolor}
\graphicspath{ {./images/} }

\usepackage{titlesec}
\titleformat{\paragraph}
{\normalfont\normalsize\bfseries}{\theparagraph}{1em}{}
\titlespacing*{\paragraph}
{0pt}{3.25ex plus 1ex minus .2ex}{1.5ex plus .2ex}

\title{Science through Machine Learning: Quantification of Poststorm Thermospheric Cooling}

\author{
 Richard J. Licata \\
  Dept. of Mechanical and Aerospace Engineering \\
  West Virginia University\\
  Morgantown, WV 26505 \\
  \texttt{rjlicata@mix.wvu.edu} \\
   \And
 Piyush M. Mehta \\
  Dept. of Mechanical and Aerospace Engineering \\
  West Virginia University \\
  Morgantown, WV \\
     \And
 Daniel R. Weimer\;\;\;\;\; \\
  Center for Space Science and Eng. Research\;\;\;\;\; \\
  Virginia Tech\;\;\;\;\; \\
  Blacksburg, VA\;\;\;\;\; \\
     \And
 Douglas P. Drob\;\;\;\;\; \\ 
  Space Science Division\;\;\;\;\; \\
  U.S. Naval Research Laboratory\;\;\;\;\; \\
  Washington, DC\;\;\;\;\; \\
     \And
   W. Kent Tobiska \\
  Space Environments Technologies \\
  Pacific Palisades, CA \\
       \And
 Jean Yoshii \\
  Space Environments Technologies \\
  Pacific Palisades, CA \\
}

\begin{document}
\maketitle
\begin{abstract}

Machine learning (ML) is often viewed as a black-box regression technique that is unable to provide considerable scientific insight. ML models are universal function approximators and -- if used correctly -- can provide scientific information related to the ground-truth dataset used for fitting. A benefit to ML over parametric models is that there are no predefined basis functions limiting the phenomena that can be modeled. In this work, we develop ML models on three datasets: the Space Environment Technologies (SET) High Accuracy Satellite Drag Model (HASDM) density database, a spatiotemporally matched dataset of outputs from the Jacchia-Bowman 2008 Empirical Thermospheric Density Model (JB2008), and an accelerometer-derived density dataset from CHAllenging Minisatellite Payload (CHAMP). These ML models are compared to the Naval Research Laboratory Mass Spectrometer and Incoherent Scatter radar (NRLMSIS 2.0) model to study the presence of post-storm cooling in the middle-thermosphere. We find that both NRLMSIS 2.0 and JB2008-ML do not account for post-storm cooling and consequently perform poorly in periods following strong geomagnetic storms (e.g. the 2003 Halloween storms). Conversely, HASDM-ML and CHAMP-ML do show evidence of post-storm cooling indicating that this phenomenon is present in the original datasets. Results show that density reductions up to 40\% can occur 1--3 days post-storm depending on location and the strength of the storm.

\end{abstract}

%%%%%%%%%%%%%%%%%%%%%%%%%%%%%%%%%%%%%%%%%%%%%%%%%%%%%%%%%%%%%%%%%%%%%%%%%%%%%%%%%%%%%%%%%%%%%%%

\section{Introduction}\label{sec:intro}

Machine learning (ML) has become increasingly prevalent across many scientific domains. This has been aided largely by improvements in computing power and the use of Graphical Processing Units (GPUs) in model training \cite{GPU_implementation}. In the space weather community, ML has been used to develop models for problems such as solar flare prediction \cite{flareML}, ionospheric scintillation detection \cite{scintML}, and geomagnetic index forecasts \cite{wingML}. However, its use is often limited to problem solving, not for investigative purposes. Further, there are rarely any studies on what the model has learned outside of determining its performance metrics. Convolutional neural networks -- specifically related to image processing -- are inherently easier to understand, as the filters (or weights) can be displayed as images and therefore interpreted \cite{alexnet}. This is not a luxury associated with ML regression models where inputs do not have visual qualities, so the weights are difficult to interpret. This motivates this work as it pertains to space weather and the thermosphere.

The thermosphere is the neutral region of the upper atmosphere consisting of many atomic and molecular species. Their relative abundance and contribution to the total mass density can vary as a function of altitude, solar activity, and space weather conditions \cite{Emmert07}. During abrupt space weather events like geomagnetic storms, the local distributions of the constituents can vary considerably causing significant changes in the total mass density \cite{storm_species}. Zesta and Oliveira \cite{heating_cooling} found that when storms become stronger, the thermosphere both heats and cools at a faster rate. Significant research has been done into a potential cause of the cooling effects, overproduction of nitric oxide (NO) and its infrared emissions.

Kockarts \cite{cooling_original} investigated the cooling impact of the thermosphere due to downward heat conduction, atomic oxygen (O), and NO during a geomagnetic storm in 1974. They found that the reduction in thermopause temperature from the introduction of NO cooling was 440 K, while the addition of O cooling only reduced the temperature by another 35 K. This topic has gained much more attention in recent years due to the NO emission data from the Sounding of the Atmosphere using Broadband Emission Radiometry (SABER) instrument %onboard the Thermosphere Ionosphere Mesosphere Energetics Dynamics (TIMED) satelllite 
\cite{SABER} and high fidelity density estimates from satellite such as CHAllenging Minisatellite Payload (CHAMP) \cite{CHAMP} and Gravity Recovery and Climate Experiment (GRACE) \cite{GRACE}.

Mlynczak et al. \cite{cooling} used SABER data during the storm periods of April 2002 and found that NO emissions were notably enhanced during this period. Lei et al \cite{overcool} considered the prominent 2003 Halloween storms to provide a comparison of SABER data to density estimates from both CHAMP and GRACE. They noted a 23--26\% maximum density depletion during the recovery phase for the satellites relative to quiet pre-storm values, and the NO cooling rates during this period remained at a high level. Knipp et al. \cite{NOstorms} examined 192 geomagnetic events to compare  NO and neutral density data from GRACE. Their data-based study suggests shock-led interplanetary coronal mass ejections result in an overproduction of NO which provides a cooling effect that compensates for the strong thermospheric expansion that occurs during these storms. The driving force behind the cooling effect is still an active area of research and other mechanisms (e.g. ionoshere-driven atomic oxygen reductions) have been proposed to explain the phenomena \cite{rebuttal,rebuttal2}. We do not attempt to confirm any driving mechanisms in this work.

Using ML, we can investigate the \textit{presence} of post-storm cooling in various datasets and which model drivers may be required to capture it. We first explain the data and models used for model development and comparison. Then, we describe the ML model development process and how we use them to examine this phenomena. We show model predictions during a prominent geomagnetic storm to motivate the importance of this work and provide a quantitative analysis on the effect of geomagnetic time history on the predicted density.

\section{Data, Models, and Methods}\label{sec:datamodelsmethods}

\subsection{Data and Models}\label{sec:data}
As a benchmark, we used the Naval Research Laboratory Mass Spectrometer and Incoherent Scatter radar (NRLMSIS 2.0) empirical thermosphere model \cite{MSIS2}, the most recent version of MSIS models dating back to the original MSIS-86 model \cite{MSIS86}. NRLMSIS 2.0 uses the \textit{ap} index to account for geomagnetic activity. The \textit{ap} index is indicative of global geomagnetic activity and has a three hour cadence. There are two \textit{ap} options when running NRLMSIS 2.0: use only the daily average (known as \textit{Ap}) and current 3-hour value, or use a time history of the index. This time history includes \textit{Ap}, current \textit{ap}, \textit{ap\textsubscript{3}}, \textit{ap\textsubscript{6}}, \textit{ap\textsubscript{9}}, \textit{ap\textsubscript{12--33}}, and \textit{ap\textsubscript{36--57}}. The single numerical subscripts refer to the value of the index that many hours prior to the epoch. The combination of two numbers in the subscript refers to the average value over that many hours prior to the epoch (e.g. \textit{ap\textsubscript{12--33}} is the average \textit{ap} value from 12 to 33 hours prior to the epoch). This nomenclature for geomagnetic drivers will be used throughout this manuscript.

We also developed machine-learned models (to be described in Section \ref{sec:ML}) based on three datasets. The first is outputs of the Jacchia-Bowman 2008 Empirical Thermospheric Density Model (JB2008) from the start of 2000 to end the of 2019 \cite{JB2008}. We evaluated JB2008 every three hours and at a fixed grid of 12,312 locations including altitude. The longitude, latitude, and altitude resolutions are 15$^{\circ}$, 10$^{\circ}$, and 25 km, respectively with the altitude ranging from 175 -- 825 km. JB2008 is driven by four solar indices/proxies. As with NRLMSIS 2.0, it uses \textit{F\textsubscript{10.7}} as a driver for solar activity. \textit{F\textsubscript{10.7}} is a valuable solar proxy that represents the 10.7 cm solar radio emission \cite{Covington}. JB2008 also uses \textit{S\textsubscript{10.7}}, \textit{M\textsubscript{10.7}}, and \textit{Y\textsubscript{10.7}}, which are explained by Tobiska et al. \cite{dev_inds} and Bowman et al. \cite{JB2008}. We do not go into their details in this work as it is not pertinent to this analysis. For geomagnetic activity, JB2008 uses both \textit{ap} and \textit{Dst}. \textit{Dst} is an index for the strength of the ring current which serves as a useful proxy for geomagnetic activity.

The second dataset is the Space Environment Technologies (SET) High Accuracy Satellite Drag Model (HASDM) density database, the first major release of outputs from the U.S. Air Force's HASDM model \cite{HASDM_SET_Data}. This HASDM dataset is spatiotemporally matched to the JB2008 outputs (same time period, time resolution, and locations). HASDM is often considered the state-of-the-art for thermospheric density modeling as it assimilates observed drag data from calibration satellites to make corrections to its background empirical model, JB2008 \cite{HASDM}.

The final dataset is the Mehta et al. \cite{CHAGRA} accelerometer-derived density estimates from CHAMP. CHAMP was in orbit from 2000 -- 2010 with a high inclination and altitude range of 300 -- 460 km. Unlike JB2008 and HASDM, this is an in-situ dataset with a much higher cadence -- 10 seconds. There are numerous other datasets that have been developed using CHAMP accelerometer data \cite{CHden1,CHden2,sutton, doorn, MarchA}, but we proceed with the described dataset due to its use in previous work \cite{QualHASDM}.

\subsection{Machine Learning}\label{sec:ML}

As the goal is to study the effects of geomagnetic time histories, we do not need to develop a surrogate model for NRLMSIS 2.0 (see Section \ref{sec:data}). We do, however, proceed with model development for the JB2008, HASDM, and CHAMP datasets. The process for JB2008 and HASDM is identical as they have the same time and space resolution. In addition, HASDM is rooted in JB2008, so we use the same inputs for both datasets. To keep model size reasonable, we leverage principal component analysis (PCA) to reduce the spatial dimensionality from 12,312 to 10 PCA coefficients. This has been demonstrated on similar datasets in the past for the development of thermospheric reduced-order models \cite{MehtaROMCal,MehtaROM,HASDM_ML}.
For information on the data preparation and PCA, the reader is referred to Licata et al. \cite{QualHASDM} where it was used to analyze the HASDM database. The major difference for CHAMP is that it is an in-situ dataset, so location is now an input as opposed to being embedded in the model output.

We first prepared the data for ML, defining the input and output data structures. JB2008 and HASDM have 26 inputs, defined in Table \ref{t:inputs}. Eight of the inputs are the four solar drivers described in Section \ref{sec:data} along with their 81-day centered averages -- marked with an "81c" subscript. These inputs are also shared with the CHAMP model. For geomagnetic activity, both JB2008 and HASDM use time histories for \textit{ap} and \textit{Dst}. The \textit{ap} time series is the one described for NRLMSIS 2.0, and the \textit{Dst} time series was determined in previous work \cite{HASDM_ML}. Since CHAMP has a much higher time resolution (1 minute), we forgo the use of these geomagnetic indices and instead use \textit{SYM-H}, which represents the longitudinally symmetric geomagnetic field disturbances \cite{sym1,sym2}. The time history was chosen to be similar to those used by the other models. We also use the Poynting flux totals in the northern and southern hemispheres (\textit{S\textsubscript{N}} and \textit{S\textsubscript{S}}) generated by the W05 electrodynamics model \cite{W05a,W05b}. The time and location inputs are described in Equations \ref{eq1} and \ref{eq2}. The general form of ($2\pi x/y$) allows for the linearly increasing inputs to be continuous about their boundaries. \textit{t\textsubscript{1}} and \textit{t\textsubscript{2}} represent annual variations, using the day of year (doy), and \textit{t\textsubscript{3}} and \textit{t\textsubscript{4}} represent diurnal variations, using universal time (UT). Equation \ref{eq2} represents the local solar time (\textit{LST}) inputs.

\begin{equation} \label{eq1}
\begin{split}
t_1=sin\left(\frac{2\pi doy}{365.25}\right),
\;\;\;\;
t_2=cos\left(\frac{2\pi doy}{365.25}\right),
\;\;\;\;
t_3=sin\left(\frac{2\pi UT}{24}\right),
\;\;\;\;
t_4=cos\left(\frac{2\pi UT}{24}\right).
\end{split}
\end{equation}

\begin{equation} \label{eq2}
\begin{split}
LST_1=sin\left(\frac{2\pi LST}{24}\right)
\;\;\;\;
LST_2=cos\left(\frac{2\pi LST}{24}\right)
\end{split}
\end{equation}

\begin{table}[htb!]
	\fontsize{10}{10}\selectfont
    \caption{List of inputs for the ML models. Note: \textit{LAT} and \textit{ALT} are the latitude and altitude at epoch, respectively.}
   \label{t:inputs}
        \centering 
   \begin{tabular}{ c | c | c } % Column formatting, 
      \hline 
          \multicolumn{3}{c}{\textbf{JB2008 / HASDM}}  \\ \hline
          \textbf{Solar} & \textbf{Geomagnetic} & \textbf{Temporal} \\ \hline 
          $F_{10}$,  $S_{10}$, & $a_{pA}$,  $a_p$,  $a_{p3}$, & $t_1$,  $t_2$ \\
          $M_{10}$,  $Y_{10}$, & $a_{p6}$,  $a_{p9}$,  $a_{p12-33}$, & $t_3$,  $t_4$ \\
          $F_{81c}$,  $S_{81c}$, & $a_{p36-57}$,  \textit{Dst}$_A$, \textit{Dst}, & \\
          $M_{81c}$,  $Y_{81c}$ & \textit{Dst}$_3$, \textit{Dst}$_6$,  \textit{Dst}$_9$, & \\
          & \textit{Dst}$_{12}$, \textit{Dst}$_{15}$, \textit{Dst}$_{18}$, \textit{Dst}$_{21}$ & \\ \hline
           \multicolumn{3}{c}{\textbf{CHAMP}} \\ \hline
          \textbf{Solar} & \textbf{Geomagnetic} & \textbf{Spatial/Temporal} \\ \hline 
          $F_{10}$,  $S_{10}$, & \textit{SYM-H}, \textit{SYM-H\textsubscript{0-3}} & $LST_1$, $LST_2$, \\
          $M_{10}$,  $Y_{10}$, & \textit{SYM-H\textsubscript{3-6}}, \textit{SYM-H\textsubscript{6-9}} & $LAT$, $ALT$, \\
          $F_{81c}$,  $S_{81c}$, & \textit{SYM-H\textsubscript{9-12}}, \textit{SYM-H\textsubscript{12-33}} & $t_1$, $t_2$, \\
          $M_{81c}$,  $Y_{81c}$ & \textit{SYM-H\textsubscript{33-57}}, $S_N$, $S_S$ & $t_3$, $t_4$\\
          \hline
   \end{tabular}
\end{table}

The outputs for JB2008 and HASDM are their respective 10 PCA coefficients, while the CHAMP outputs are the local density estimates. With the inputs and outputs set up in an ML format, we can determine an architecture for each dataset using Keras Tuner, an automated tool for hyperparameter searching \cite{KT}. The tuner is provided a hyperparameter space (e.g. choices for the number of layers, neurons, activation functions, and optimizers) and trains three models with random weight initialization for 100 architectures or trials. The first 25 architectures are randomly selected from the search space. After the random search, the tuner chooses all subsequent architectures using a Bayesian optimization scheme attempting to minimize the loss on the validation set. The ten best models are evaluated on the training and validation sets to determine the final model considering the lowest errors. The model development for CHAMP, HASDM, and subsequently JB2008 is outlined by Licata and Mehta \cite{UQtech} with the only difference being the \textit{SYM-H} time series inputs for CHAMP. We provide additional details on model development in Appendices \hyperref[sec:appA]{A} and \hyperref[sec:appB]{B}.

\subsection{Storm Example}

To motivate the work, we evaluated NRLMSIS 2.0 and the ML models from the JB2008, HASDM, and CHAMP datasets during the 2003 Halloween storms. The latter models will be referred to as JB2008-ML, HASDM-ML, and CHAMP-ML, respectively. We used the time series \textit{ap} flag when running NRLMSIS 2.0 in this work. The four models were provided the true drivers for the six day period from October 28 -- November 3, 2003 and were compared to the Mehta et al. \cite{CHAGRA} CHAMP estimates. For NRLMSIS 2.0 and CHAMP-ML, the predictions were made with the same time cadence and at the specific locations of the satellite, negating the need for further processing.  For JB2008-ML and HASDM-ML, we made predictions at the 3-hour intervals used by the original models. Those 3D density grids were then interpolated in space and time in log-scale to the locations of CHAMP. The final step is to take a running average of the along-orbit densities over the 92.5 minute orbital period to obtain orbit-averaged densities. This allows us to visualize the general density along the orbit during the storm period (Figure \ref{f:storm}).

\subsection{Time Lag Study} \label{sec:timelagstudy}

As discussed in Section \ref{sec:intro}, cooling mechanisms often cause post-storm densities to be anomalously low. For this storm in particular, Lei et al. \cite{overcool} noted nearly a 25\% decrease in post-storm densities relative to pre-storm levels. In an effort to quantify this mechanism within the original models/datasets, we varied the time histories for \textit{ap} or \textit{SYM-H} independently within the models at four locations listed in Table \ref{t:lag_info}. Table \ref{t:lag_info} also contains the geomagnetic indices held constant in each model while either \textit{ap} or \textit{SYM-H} were changed. All cases were at a constant solar activity with drivers set to 120. The time inputs were at 0 hours UT and represent the fall equinox (doy = 264), so there are no effects from Earth's tilt.

\begin{table}[htb!]
	\fontsize{10}{10}\selectfont
    \caption{Information for the time lag study. For clarification, LAT is latitude and \textit{S} refers to both \textit{S\textsubscript{N}} and \textit{S\textsubscript{S}}.}
   \label{t:lag_info}
        \centering 
   \begin{tabular}{ c | c | c | c } % Column formatting, 
      \hline 
        \multicolumn{4}{c}{\textbf{Locations}} \\ \hline
        \textbf{Night Equator} & \textbf{Day Equator} & \textbf{Night Pole} & \textbf{Day Pole} \\ \hline
        LST = 2 hrs, LAT = 0$^{\circ}$ & LST = 14 hrs, LAT = 0$^{\circ}$ & LST = 2 hrs, LAT = 80$^{\circ}$ & LST = 14 hrs, LAT = 80$^{\circ}$ \\ \hline
        \multicolumn{4}{c}{\textbf{Constant Inputs}} \\ \hline
        \textbf{NRLMSIS 2.0} & \textbf{JB2008-ML} & \textbf{HASDM-ML} & \textbf{CHAMP-ML} \\ \hline
        \textit{ap} = 56 & \textit{ap} = 56, \textit{Dst} = -50 & \textit{ap} = 56, \textit{Dst} = -50 & \textit{SYM-H} = -50, \textit{S} = 200  \\ \hline
   \end{tabular}
\end{table}

The models were run with the aforementioned inputs held constant and each of the time-series geomagnetic indices increased in magnitude independently. The three models using \textit{ap} step through the 28 discrete values between 0 and 400. For CHAMP-ML, \textit{SYM-H} decreased from 0 to -250 in increments of 5. We then computed the density ratios for each time series index with respect to the density when the index is set to 0. We plot all the curves generated for each location/model (Figure \ref{f:time_hist})% and show a supplementary table with the density ratios at \textit{ap} = 207 and \textit{ap} = 400 or \textit{SYM-H} = -125 and \textit{SYM-H} = -250
.

\section{Results and Discussion}

We first show the error statistics for the three ML models developed in Table \ref{t:stats}. These were computed with respect to the original datasets. Training data is used to fit the model, validation data is used to determine the best model, and the independent test set used to determine performance; details are in Appendix \hyperref[sec:appA]{A}.

\begin{table}[htb!]
	\fontsize{10}{10}\selectfont
    \caption{Mean absolute error on the training, validation, and test sets: 100\% abs(true-predicted) / true.}
   \label{t:stats}
        \centering 
   \begin{tabular}{ c | c | c | c } % Column formatting, 
      \hline 
        \textbf{Model} & \textbf{Training} & \textbf{Validation} & \textbf{Test} \\ \hline
        \textbf{JB2008-ML} & 5.28\% & 6.03\% & 6.63\% \\ \hline
        \textbf{HASDM-ML} & 9.13\% & 10.46\% & 10.39\% \\ \hline
        \textbf{CHAMP-ML} & 10.97\% & 11.60\% & 11.57\% \\ \hline
   \end{tabular}
\end{table}

\begin{figure}[htb!]
	\centering
	\small
	\includegraphics[width=\textwidth]{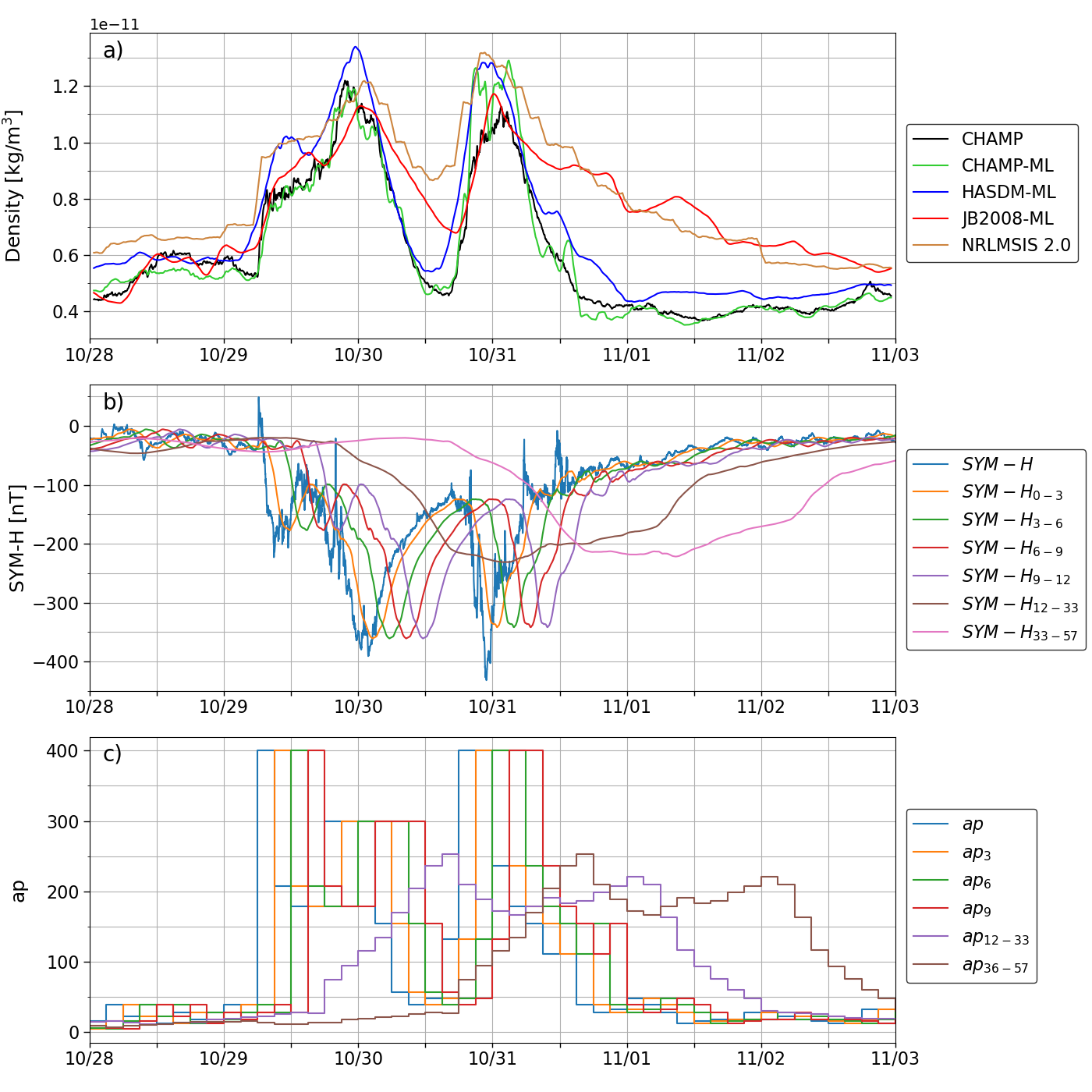}
    \caption{Orbit-average density for NRLMSIS 2.0, JB2008-ML, HASDM-ML, CHAMP-ML, and CHAMP (a) and the associated \textit{SYM-H} (b) and \textit{ap} (c) time-series inputs.}
	\label{f:storm}
\end{figure}

Table \ref{t:stats} shows that JB2008-ML undoubtedly has the lowest errors, but it is worth noting that it is also the most generalized dataset of the three. The SET HASDM density database contains evidence of more complicated processes and its PCA coefficients are more difficult to model as a result \cite{QualHASDM}. The CHAMP-ML errors are 1--2\% higher than those of HASDM-ML. CHAMP-ML is the only ML model in this work with location as a predictor. To both visualize the model performance in an operational setting and motivate the remainder of the work, we show the orbit-averaged densities for the three ML models and NRLMSIS 2.0 compared to the Mehta et al. \cite{CHAGRA} CHAMP densities for the 2003 Halloween storms in Figure \ref{f:storm}.

Figure \ref{f:storm} (a) shows that all models match the timing observed by CHAMP during both storms (10/29--10/30 and 10/30--10/31). NRLMSIS 2.0 has a tendency to overpredict density throughout this 6-day period, most notably between the two storms and in the recovery phase (11/02--11/03). JB2008-ML exhibits similar behavior although it is closer to matching the contraction of the atmosphere between the storms. While both of these models use time-histories of \textit{ap}, they do not portray any evidence of post-storm cooling. In contrast, HASDM-ML and CHAMP-ML both show significant decreases in density both between and after the storms. 

\begin{figure}[htb!]
	\centering
	\small
	\includegraphics[width=1.00\textwidth]{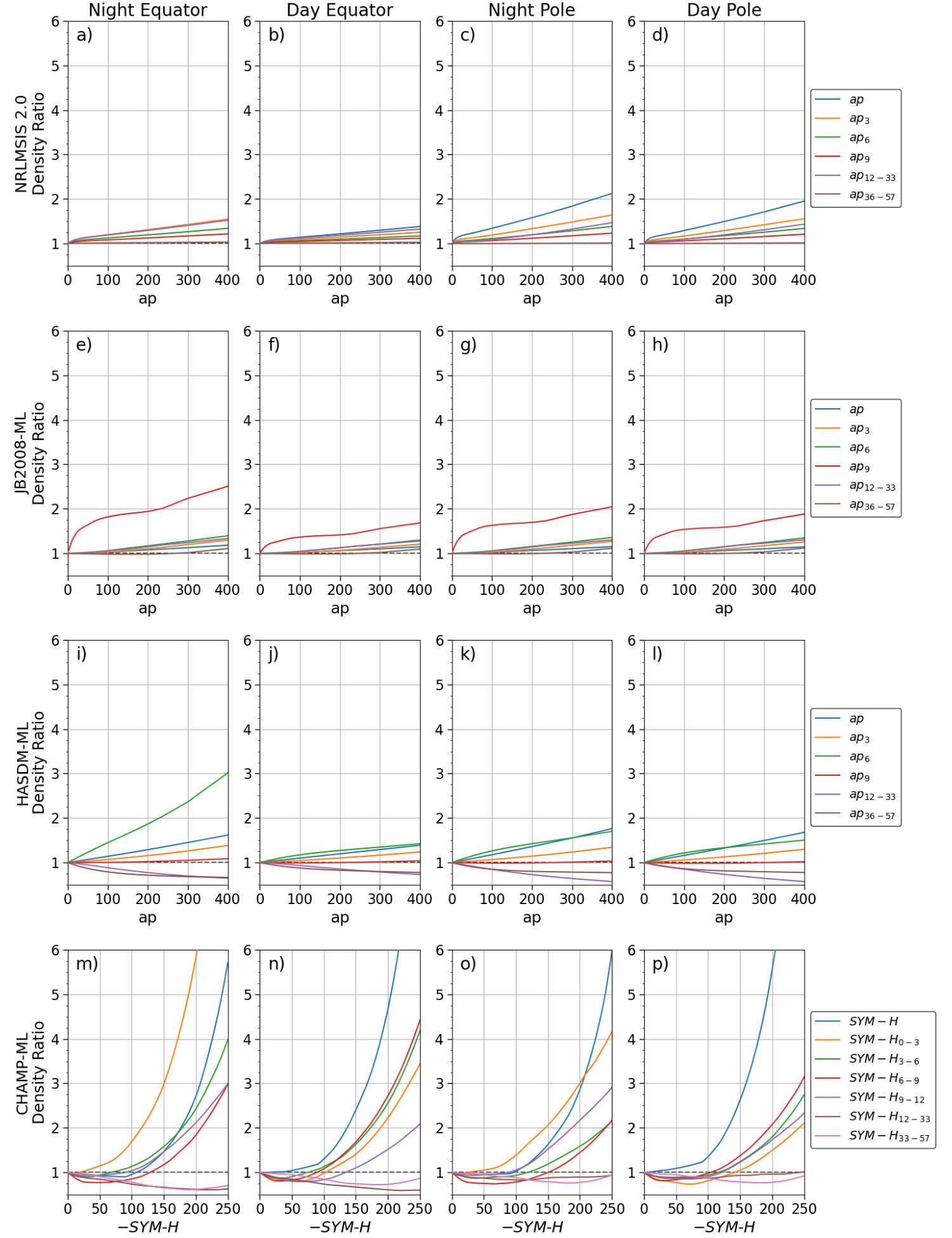}
    \caption{Density ratios for the four locations and four models, as described in Section \ref{sec:timelagstudy}.}
	\label{f:time_hist}
\end{figure}

Figure \ref{f:storm} (b) and (c) show \textit{SYM-H} and \textit{ap} time histories, respectively. The increased temporal resolution for \textit{SYM-H} is very evident, and the first four averages can inform CHAMP-ML of the recent magnetic disturbances. The last two time history inputs (\textit{SYM-H\textsubscript{12-33}} and \textit{SYM-H\textsubscript{33-57}}) represent longer-term information with less variation. Immediately following the second storm (around 0600 UTC on 10/31), the last two time history inputs have large magnitudes while the more recent inputs no longer signify a storm. At the same time, the density predicted by CHAMP-ML and observed by the satellite drop abruptly. This behavior reinforces observations of Zesta and Oliveira \cite{heating_cooling}. The \textit{ap} time history is valuable to the other three models, following similar trends to panel (b) but are much more coarse. The results from the time-lag study (described in Section \ref{sec:timelagstudy}) are displayed in Figure \ref{f:time_hist}.% and Table \ref{t:ratios}.

% \clearpage

Figure \ref{f:time_hist} % and Table \ref{t:ratios} are 
is informative into what historical information is most important to represent the original data source -- JB2008 output, SET HASDM density database, and CHAMP density estimates. NRLMSIS 2.0 is used here as a baseline due to its wide use in the field and use of historical geomagnetic information. There is a fairly linear relationship between the different \textit{ap} values and density for NRLMSIS 2.0. In most cases, it considers the most recent \textit{ap} to be most important and the least recent \textit{ap} to be the least important. There is almost a perfect decay of slopes as it considers information from further in the past. At no point does the density ratio at the four locations drop below 1.00, which represents lower density than the baseline, or \textit{ap\textsubscript{x}} = 0 where \textit{x} represents a given time-lag or lack thereof.

For JB2008-ML there is virtually no evidence of post-storm cooling being present in the dataset. With the exception of the \textit{ap\textsubscript{9}} curves, there is a fairly linear dependence between \textit{ap} and density. Interestingly, JB2008-ML indicates that the strongest relationship between \textit{ap} and density has a 9-hour delay. The \textit{ap\textsubscript{9}} curves are nonlinear for \textit{ap} < 100 and quite linear for \textit{ap} > 100. While there are values for JB2008-ML in %Table \ref{t:ratios} 
Figure \ref{f:time_hist} that are less than 1.00, they are at most showing a 2\% decrease and only at the equatorial locations.

HASDM-ML has a near-linear relationship with \textit{ap}, but there is considerable evidence of post-storm cooling seen in % both 
Figure \ref{f:time_hist}% and Table \ref{t:ratios}
. At each of the four locations, \textit{ap\textsubscript{12-33}} and \textit{ap\textsubscript{36-57}} have a inverse relationship with density. At the two high-latitude locations, \textit{ap\textsubscript{12-33}} causes the lowest density ratios while \textit{ap\textsubscript{36-57}} causes the lowest density ratios at the equator. This may be a result of the time-delay of the density response at low latitudes relative to the auroral region. In contrast to JB2008-ML panels (e)-(h), HASDM-ML has a strong positive relationship between \textit{ap\textsubscript{6}} and density with little impact from \textit{ap\textsubscript{9}}. At the highest levels of activity (\textit{ap} > 300), the current \textit{ap} value drives the strongest increase in density at the poles. 

CHAMP-ML displays a highly nonlinear relationship between \textit{SYM-H} and density. At each location, the relative importance of each input can change significantly; the maximum density ratio for \textit{SYM-H\textsubscript{0-3}} is 10.25 at the nightside equator while it is only 2.10 at the dayside pole. There is strong evidence of post-storm cooling in the CHAMP dataset, highlighted by the array of historical \textit{SYM-H} drivers causing density ratios below 1.00. The least recent \textit{SYM-H} averages have their strongest inverse relationship with density at the equatorial locations while other historical indices demonstrate low density ratios at the polar location. %In Table \ref{t:ratios}, the 
The CHAMP-ML density ratios drop as low as 0.59 %0.61 
and rise as high as 12.34 indicating a more complex relationship between geomagnetic activity and density compared to the other three models in this analysis.

\section{Summary}

In this work, we demonstrated the use of machine-learned models to investigate thermospheric post-storm cooling. We trained ML models on three datasets, JB2008 outputs, the SET HASDM density database, and the Mehta et al. \cite{CHAGRA} CHAMP density estimates, to conduct this assessment and used NRLMSIS 2.0 for comparison. All models developed were provided a recent time history (up to 57 hours) of geomagnetic drivers to see if the data suggests that there is evidence of post-storm cooling; the models would need to see that previous geomagnetic drivers indicate a storm of a given strength has recently occurred. Using the 2003 Halloween storms as an example (Figure \ref{f:storm}), we showed that both NRLMSIS 2.0 and JB2008-ML do not match the sudden cooling seen between and after the two storms by the CHAMP accelerometer. Meanwhile, HASDM-ML and CHAMP-ML both model the general density trends of this storm and display attributes of an abruptly cooled thermosphere.

When considering a historical event, other factors play a role in how the thermosphere behaves. Therefore, we isolated the internal model formulation only as it pertains to recent magnetic perturbations. By holding all model drivers constant and only vary a single geomagnetic driver at a time: \textit{ap} for NRLMSIS 2.0, JB2008-ML and HASDM-ML and \textit{SYM-H} for CHAMP-ML. This showed that NRLMSIS 2.0 and JB2008-ML both did not exhibit any cooling effects as the historical \textit{ap} values were raised, which would indicate a strong storm had recently taken place. In fact, the most important historical driver to JB2008-ML was the 9-hour prior \textit{ap} which resulted in density ratios nearly twice that of any other driver.

HASDM-ML was most strongly driven by the current and 6-hour prior \textit{ap} for thermospheric expansion while increases in \textit{ap\textsubscript{12-33}} and \textit{ap\textsubscript{36-57}} resulted in densities as low as 57\% of the baseline magnitude% (Table \ref{t:ratios})
. CHAMP-ML was the only model to indicate a nonlinear relationship between density and geomagnetic activity. Depending on the location, \textit{SYM-H} or \textit{SYM-H\textsubscript{0-3}} drove the largest density ratios, significantly more than any other model. In terms of cooling, CHAMP-ML showed that at \textit{SYM-H} > -100 nT, many of the recent drivers caused a density ratio less than 1.00. As the index was made more negative, the least recent drivers caused the lowest density ratios, particularly at low latitudes. 

While the way the historical geomagnetic indices were varied may be nonphysical, this evaluated what relationship these ML regression models learned from their respective datasets. The "hidden" internal structure of the ML regression models are difficult to examine, but an analysis like this can inform us what the overall formulation the model has with respect to its drivers. This method can be used to gain a better understanding of the model and to investigate complex space weather datasets.

\section*{Data Statement}

Requests can be submitted for full access to the SET HASDM density database at \url{https://spacewx.com/hasdm/} and all reasonable requests for scientific research are accepted as explained in the rules of road document on the website. The historical space weather indices used in this study can be found at the following links: \textit{F\textsubscript{10.7}}: \url{https://www.spaceweather.gc.ca/forecast-prevision/solar-solaire/solarflux/sx-en.php}, \textit{ap}: \url{https://doi.org/10.5880/Kp.0001}, \textit{Dst}: \url{http://wdc.kugi.kyoto-u.ac.jp/dstdir/}, and \textit{SYM-H}: \url{http://wdc.kugi.kyoto-u.ac.jp/aeasy/index.html}. The remaining solar indices and proxies can be found at \url{https://spacewx.com/jb2008/} in the SOLFSMY.TXT file. The Weimer \cite{PF} Pointing flux data can be accessed at \url{https://doi.org/10.5281/zenodo.3525166}. Free, one-time only registration is required to access the historical data while nowcasts and forecasts are provided by SET as a data service from data@spacewx.com. CHAMP and GRACE position data were obtained from the measurements presented by Mehta et al. \cite{CHAGRA} at \url{http://tinyurl.com/densitysets}.

\section*{Acknowledgements}

PMM gratefully acknowledges support under NSF CAREER award \#2140204 and NSF ANSERS award \#2149747 (subaward to WVU from Rutgers). All authors acknowledge support by NASA grant \#80NSSC20K1362 to Virginia Tech under the Space Weather Operations 2 Research Program, with subcontracts to WVU and SET. Douglas Drob was supported by NASA interagency agreement 80HQTR20T0081 with the Naval Research Laboratory. We would like to thank Space Weather Canada for providing and maintaining solar radio emission data, GFZ Potsdam for supplying \textit{ap} archives, and the World Data Center for Geomagnetism in Kyoto for providing \textit{Dst} and \textit{SYM-H} data. The authors would like to acknowledge DLR for their work on the CHAMP mission along with GFZ Potsdam for managing the data.

\bibliographystyle{ieeetr}  
\bibliography{references}

\begin{thebibliography}{10}

\bibitem{GPU_implementation}
K.~Chellapilla, S.~Puri, and P.~Simard, ``{High Performance Convolutional
  Neural Networks for Document Processing},'' in {\em {Tenth International
  Workshop on Frontiers in Handwriting Recognition}} (G.~Lorette, ed.), (La
  Baule (France)), {Universit{\'e} de Rennes 1}, {Suvisoft}, 2006.

\bibitem{flareML}
K.~Florios, I.~Kontogiannis, S.-H. Park, J.~A. Guerra, F.~Benvenuto, D.~S.
  Bloomfield, and M.~K. Georgoulis, ``Forecasting solar flares using
  magnetogram-based predictors and machine learning,'' {\em Solar Physics},
  vol.~293, no.~2, pp.~1--42, 2018.

\bibitem{scintML}
Y.~Jiao, J.~J. Hall, and Y.~T. Morton, ``"automatic equatorial gps amplitude
  scintillation detection using a machine learning algorithm",'' {\em IEEE
  Transactions on Aerospace and Electronic Systems}, vol.~53, no.~1,
  pp.~405--418, 2017.

\bibitem{wingML}
S.~Wing, J.~R. Johnson, J.~Jen, C.-I. Meng, D.~G. Sibeck, K.~Bechtold,
  J.~Freeman, K.~Costello, M.~Balikhin, and K.~Takahashi, ``Kp forecast
  models,'' {\em Journal of Geophysical Research: Space Physics}, vol.~110,
  no.~A4, 2005.

\bibitem{alexnet}
A.~Krizhevsky, I.~Sutskever, and G.~E. Hinton, ``"imagenet classification with
  deep convolutional neural networks",'' in {\em "Advances in Neural
  Information Processing Systems"} (F.~Pereira, C.~J.~C. Burges, L.~Bottou, and
  K.~Q. Weinberger, eds.), vol.~25, Curran Associates, Inc., 2012.

\bibitem{Emmert07}
J.~Emmert, ``{Thermospheric mass density: A review},'' {\em Advances in Space
  Research}, vol.~56, 06 2015.

\bibitem{storm_species}
T.~J. Fuller-Rowell, M.~V. Codrescu, R.~G. Roble, and A.~D. Richmond, {\em {How
  Does the Thermosphere and Ionosphere React to a Geomagnetic Storm?}},
  pp.~203--225.
\newblock American Geophysical Union (AGU), 1997.

\bibitem{heating_cooling}
E.~Zesta and D.~M. Oliveira, ``Thermospheric heating and cooling times during
  geomagnetic storms, including extreme events,'' {\em Geophysical Research
  Letters}, vol.~46, no.~22, pp.~12739--12746, 2019.

\bibitem{cooling_original}
G.~Kockarts, ``Nitric oxide cooling in the terrestrial thermosphere,'' {\em
  Geophysical Research Letters}, vol.~7, no.~2, pp.~137--140, 1980.

\bibitem{SABER}
J.~M.~I. Russell, M.~G. Mlynczak, L.~L. Gordley, J.~Tansock, and R.~Esplin,
  ``{An Overview of the SABER Experiment and Preliminary Calibration
  Results},'' {\em Space Dynamics Lab Publications}, vol.~114, 1999.

\bibitem{CHAMP}
C.~Reigber, H.~L{\"u}hr, and P.~Schwintzer, ``Champ mission status,'' {\em
  Advances in space research}, vol.~30, no.~2, pp.~129--134, 2002.

\bibitem{GRACE}
S.~Bettadpur, ``{Gravity Recovery and Climate Experiment: Product Specification
  Document},'' {\em {GRACE 327-720, CSR-GR-03-02}}, 2012.
\newblock Cent. for Space Res., The Univ. of Texas, Austin, TX,
  \url{https://podaac.jpl.nasa.gov/GRACE}.

\bibitem{cooling}
M.~Mlynczak, F.~J. Martin-Torres, J.~Russell, K.~Beaumont, S.~Jacobson,
  J.~Kozyra, M.~Lopez-Puertas, B.~Funke, C.~Mertens, L.~Gordley, R.~Picard,
  J.~Winick, P.~Wintersteiner, and L.~Paxton, ``{The natural thermostat of
  nitric oxide emission at 5.3 $\mu m$ in the thermosphere observed during the
  solar storms of April 2002},'' {\em Geophysical Research Letters}, vol.~30,
  no.~21, 2003.

\bibitem{overcool}
J.~Lei, A.~G. Burns, J.~P. Thayer, W.~Wang, M.~G. Mlynczak, L.~A. Hunt, X.~Dou,
  and E.~Sutton, ``{Overcooling in the upper thermosphere during the recovery
  phase of the 2003 October storms},'' {\em Journal of Geophysical Research:
  Space Physics}, vol.~117, no.~A3, 2012.

\bibitem{NOstorms}
D.~J. Knipp, D.~V. Pette, L.~M. Kilcommons, T.~L. Isaacs, A.~A. Cruz, M.~G.
  Mlynczak, L.~A. Hunt, and C.~Y. Lin, ``Thermospheric nitric oxide response to
  shock-led storms,'' {\em Space Weather}, vol.~15, no.~2, pp.~325--342, 2017.

\bibitem{rebuttal}
A.~V. Mikhailov and L.~Perrone, ``{Poststorm Thermospheric NO Overcooling?},''
  {\em Journal of Geophysical Research: Space Physics}, vol.~125, no.~1,
  p.~e2019JA027122, 2020.

\bibitem{rebuttal2}
J.~Lei, W.~Wang, A.~G. Burns, S.-R. Zhang, and T.~Dang, ``{Comments on
  “Poststorm Thermospheric NO Overcooling?” by Mikhailov and Perrone
  (2020)},'' {\em Journal of Geophysical Research: Space Physics}, vol.~126,
  no.~4, p.~e2020JA027992, 2021.

\bibitem{MSIS2}
J.~T. Emmert, D.~P. Drob, J.~M. Picone, D.~E. Siskind, M.~Jones~Jr., M.~G.
  Mlynczak, P.~F. Bernath, X.~Chu, E.~Doornbos, B.~Funke, L.~P. Goncharenko,
  M.~E. Hervig, M.~J. Schwartz, P.~E. Sheese, F.~Vargas, B.~P. Williams, and
  T.~Yuan, ``{NRLMSIS 2.0: A Whole-Atmosphere Empirical Model of Temperature
  and Neutral Species Densities},'' {\em Earth and Space Science}, vol.~8,
  no.~3, p.~e2020EA001321, 2021.
\newblock e2020EA001321 2020EA001321.

\bibitem{MSIS86}
A.~E. Hedin, ``{MSIS-86 Thermospheric Model},'' {\em Journal of Geophysical
  Research: Space Physics}, vol.~92, no.~A5, pp.~4649--4662, 1987.

\bibitem{JB2008}
B.~Bowman, W.~K. Tobiska, F.~Marcos, C.~Huang, C.~Lin, and W.~Burke, ``{A New
  Empirical Thermospheric Density Model JB2008 Using New Solar and Geomagnetic
  Indices},'' in {\em AIAA/AAS Astrodynamics Specialist Conference}, AIAA
  2008-6438, 2008.
\newblock \url{https://arc.aiaa.org/doi/abs/10.2514/6.2008-6438}.

\bibitem{Covington}
A.~E. {Covington}, ``{Solar Noise Observations on 10.7 Centimeters},'' {\em
  Proceedings of the IRE}, vol.~36, no.~4, pp.~454--457, 1948.

\bibitem{dev_inds}
W.~K. Tobiska, S.~D. Bouwer, and B.~R. Bowman, ``The development of new solar
  indices for use in thermospheric density modeling,'' {\em Journal of
  Atmospheric and Solar-Terrestrial Physics}, vol.~70, no.~5, pp.~803--819,
  2008.

\bibitem{HASDM_SET_Data}
W.~K. Tobiska, B.~R. Bowman, D.~Bouwer, A.~Cruz, K.~Wahl, M.~Pilinski, P.~M.
  Mehta, and R.~J. Licata, ``{The SET HASDM density database},'' {\em Space
  Weather}, p.~e2020SW002682, 2021.

\bibitem{HASDM}
M.~F. Storz, B.~R. Bowman, M.~J.~I. Branson, S.~J. Casali, and W.~K. Tobiska,
  ``High accuracy satellite drag model (hasdm),'' {\em Advances in Space
  Research}, vol.~36, no.~12, pp.~2497--2505, 2005.

\bibitem{CHAGRA}
P.~M. Mehta, A.~C. Walker, E.~K. Sutton, and H.~C. Godinez, ``{New density
  estimates derived using accelerometers on board the CHAMP and GRACE
  satellites},'' {\em Space Weather}, vol.~15, no.~4, pp.~558--576, 2017.

\bibitem{CHden1}
S.~Bruinsma and R.~Biancale, ``{Total Densities Derived from Accelerometer
  Data},'' {\em Journal of Spacecraft and Rockets}, vol.~40, no.~2,
  pp.~230--236, 2003.

\bibitem{CHden2}
H.~Liu, H.~L\"uhr, V.~Henize, and W.~Köhler, ``{Global distribution of the
  thermospheric total mass density derived from CHAMP},'' {\em Journal of
  Geophysical Research: Space Physics}, vol.~110, no.~A4, 2005.

\bibitem{sutton}
E.~K. {Sutton}, {\em {Effects of solar disturbances on the thermosphere
  densities and winds from CHAMP and GRACE satellite accelerometer data}}.
\newblock PhD thesis, University of Colorado at Boulder, Oct. 2008.
\newblock \url{https://ui.adsabs.harvard.edu/abs/2008PhDT........87S}.

\bibitem{doorn}
E.~Doornbos, {\em {Producing Density and Crosswind Data from Satellite Dynamics
  Observations}}, pp.~91--126.
\newblock Berlin, Heidelberg: Springer Berlin Heidelberg, 2012.

\bibitem{MarchA}
G.~March, E.~Doornbos, and P.~Visser, ``{High-fidelity geometry models for
  improving the consistency of CHAMP, GRACE, GOCE and Swarm thermospheric
  density data sets},'' {\em Advances in Space Research}, vol.~63, no.~1,
  pp.~213--238, 2019.

\bibitem{QualHASDM}
R.~J. Licata, P.~M. Mehta, W.~K. Tobiska, B.~R. Bowman, and M.~D. Pilinski,
  ``{Qualitative and Quantitative Assessment of the SET HASDM Database},'' {\em
  Space Weather}, vol.~19, no.~8, p.~e2021SW002798, 2021.

\bibitem{MehtaROMCal}
P.~M. Mehta and R.~Linares, ``A methodology for reduced order modeling and
  calibration of the upper atmosphere,'' {\em Space Weather}, vol.~15, no.~10,
  pp.~1270--1287, 2017.

\bibitem{MehtaROM}
P.~M. Mehta, R.~Linares, and E.~K. Sutton, ``{A Quasi-Physical Dynamic Reduced
  Order Model for Thermospheric Mass Density via Hermitian Space-Dynamic Mode
  Decomposition},'' {\em Space Weather}, vol.~16, no.~5, pp.~569--588, 2018.

\bibitem{HASDM_ML}
R.~J. Licata, P.~M. Mehta, W.~K. Tobiska, and S.~Huzurbazar, ``{Machine-Learned
  HASDM Thermospheric Mass Density Model with Uncertainty Quantification},''
  {\em Space Weather}, vol.~20, p.~e2021SW002915, 2022.
\newblock
  \url{https://agupubs.onlinelibrary.wiley.com/doi/abs/10.1029/2021SW002915}.

\bibitem{sym1}
T.~Iyemori, ``{Storm-Time Magnetospheric Currents Inferred from Mid-Latitude
  Geomagnetic Field Variations},'' {\em Journal of geomagnetism and
  geoelectricity}, vol.~42, no.~11, pp.~1249--1265, 1990.

\bibitem{sym2}
F.~Siciliano, G.~Consolini, R.~Tozzi, M.~Gentili, F.~Giannattasio, and
  P.~De~Michelis, ``{Forecasting SYM-H Index: A Comparison Between Long
  Short-Term Memory and Convolutional Neural Networks},'' {\em Space Weather},
  vol.~19, no.~2, p.~e2020SW002589, 2021.

\bibitem{W05a}
D.~R. Weimer, ``{Improved ionospheric electrodynamic models and application to
  calculating Joule heating rates},'' {\em Journal of Geophysical Research:
  Space Physics}, vol.~110, no.~A5, 2005.

\bibitem{W05b}
D.~R. Weimer, ``Predicting surface geomagnetic variations using ionospheric
  electrodynamic models,'' {\em Journal of Geophysical Research: Space
  Physics}, vol.~110, no.~A12, 2005.

\bibitem{KT}
T.~O'Malley, E.~Bursztein, J.~Long, F.~Chollet, H.~Jin, L.~Invernizzi, {\em
  et~al.}, ``{Keras Tuner}.'' \url{https://github.com/keras-team/keras-tuner},
  2019.

\bibitem{UQtech}
R.~J. Licata and P.~M. Mehta, ``{Uncertainty Quantification Techniques for
  Space Weather Modeling: Thermospheric Density Application},'' 2022.

\bibitem{PF}
D.~Weimer, ``{Improving Neutral Density Predictions Using Exospheric
  Temperatures Calculated on a Geodesic, Polyhedral Grid [Data set]},'' Nov.
  2019.
\newblock \url{https://doi.org/10.5281/zenodo.3525166}.

\end{thebibliography}

\section*{Appendix A: Data Splitting} \label{sec:appA}
\setcounter{table}{0}
\renewcommand{\thetable}{A\arabic{table}}

When developing a ML model, one must carefully consider how to split the data into training, validation, and test sets. These models can have thousands or even millions of parameters, so it is critical to ensure the analysis is valid and that the model has not just memorized the training data. The goal is to provide enough diverse data in the training set for the model to learn the proper relationship between the inputs and outputs. The validation and test sets are chosen to also contain diverse conditions. The validation set is used to determine the model performance -- during and after training -- on data not used for fitting. However, to choose the best version of the model based on its training and validation performance, there must be another independent dataset for model evaluation; this is called the test set.

For JB2008-ML and HASDM-ML, the data splitting is straightforward. Looking at the solar and geomagnetic drivers, we split the data by years trying to place various portions of the solar cycle in each set. The breakdown is displayed in Table \ref{t:TVT_JBHA}. This results in 12 years being used for training (60\%), 4 years being used for validation (20\%), and 4 years being used for test (20\%).

\begin{table}[htb!]
	\fontsize{10}{10}\selectfont
    \caption{Training, validation, and test split for JB2008-ML and HASDM-ML by year.}
   \label{t:TVT_JBHA}
        \centering 
   \begin{tabular}{ c | c | c | c } % Column formatting, 
      \hline 
        \textbf{Set} & \textbf{Training} & \textbf{Validation} & \textbf{Test} \\ \hline
        \multirow{3}{*}{\textbf{Years}} & 2000, 2001, 2003, 2004, & 2005, 2008, 2014, 2017 & 2002, 2006, 2010, 2018  \\ 
        & 2007, 2009, 2011, 2012, & & \\
        & 2013, 2015, 2016, 2019 & & \\ \hline
   \end{tabular}
\end{table}

For CHAMP-ML, the combination of having location as an input and not having a full solar cycle of data makes the data splitting more complex. Using year-long periods (as in Table \ref{t:TVT_JBHA}) leaves out an entire range of possible altitudes and solar drivers from training. We work around this by repeating the following scheme: 8 consecutive weeks for training, 1 week for validation, and 1 week for test. Even though there are only two weeks between training segments, the 10-second cadence of the measurements creates a gap of over 120,000 samples, ensuring the model is not just simply interpolating for the validation and test sets. While only 60\% of data is in the JB2008-ML and HASDM-ML training sets, the spatiotemporally limited CHAMP dataset requires more training data to prevent overfitting.

\section*{Appendix B: Additional Model Development Details} \label{sec:appB}
\setcounter{table}{0}
\renewcommand{\thetable}{B\arabic{table}}
\setcounter{figure}{0}
\renewcommand{\thefigure}{B\arabic{figure}}

\subsection*{Keras Tuner}

As discussed in Section \ref{sec:ML}, we use Keras Tuner to identify an architecture for each dataset. The hyperparameter space we defined for the tuners (one for each dataset) is displayed in Table \ref{t:hyp}.

\begin{table}[htb!]
	\fontsize{10}{10}\selectfont
    \caption{Hyperparameter search space for the three ML models.}
   \label{t:hyp}
        \centering 
   \begin{tabular}{ c | c } % Column formatting, 
      \hline 
            \textbf{Parameter} & \textbf{Choices} \\ \hline 
            \textit{Number of Hidden Layers} & 1 -- 10\\ \hline 
            \textit{Neurons} & min = 64, max = 1024, step = 4\\ \hline 
            \multirow{2}{*}{\textit{Activations}} & relu, softplus, tanh, sigmoid, \\ 
            & softsign, selu, elu, linear\\ \hline
            \textit{Dropout} & min = 0.10, max = 0.60, step = 0.01\\ \hline
            \multirow{2}{*}{\textit{Optimizer}} & RMSprop, Adam, Nadam,  \\ 
            & Adadelta, Adagrad \\
      \hline
   \end{tabular}
\end{table}

For JB2008-ML and HASDM-ML, the tuner is provided the entire training and validation sets due to the relatively small size of the datasets. Therefore, once the tuner returns the best models, each trained for 100 training iterations or epochs, there is no need for further training. In fact, training after tuning did not yield improved models in these two cases. The CHAMP-ML training and validation sets have over 20 million and 2 million samples, respectively. Therefore, we only supply the tuner with 1 million random samples from the training set and 200,000 random samples from the validation set. This provides the tuner with ample information to determine an adequate architecture from which we can develop the full model. This full model will use the full training and validation sets.

\subsection*{Loss Function and Custom Layer}

The loss function plays an important role in the tuning and training process. It is the metric that the fitting function tries to minimize or maximize through the weight updates. For this study, a mean square error (MSE) loss function would suffice since the goal is to develop a model with minimal error with respect to the respective dataset. However, we used models from previous work, Licata et al \cite{UQtech}, which used the negative logarithm of predictive density (NLPD) loss function,

\begin{equation} \label{eqNLPD}
\begin{split}
NLPD(y,\mu,\sigma) = \frac{\left(y-\mu\right)^2}{2\sigma^2} + \frac{ln(\sigma^2)}{2} + \frac{ln(2\pi)}{2}
\end{split}
\end{equation}

where $y$, $\mu$, and $\sigma$ are the ground truth, mean prediction, and standard deviation of the prediction, respectively. In practice, the factor of 1/2 and the last term can be removed. There are no ground truth values for $\sigma$ for these datasets, but we are able to directly predict it since it is a standalone term in Equation \ref{eqNLPD}. This is accomplished by using twice the output neurons as there are outputs where they represent the mean and standard deviation of the model uncertainty. The output neurons representing the mean use a linear activation function, while those representing the standard deviation use the "softplus" activation function. The softplus activation and its derivative, the sigmoid function, are defined as,

\begin{equation} \label{eqAct}
\begin{split}
f(x) = ln(1+e^x) \;\;\;\;\;\;\;\;\;\;\; f'(x) = \frac{e^x}{1+e^x}
\end{split}
\end{equation}

with $x$ being the weighted sum of the neurons inputs. Licata et al. \cite{HASDM_ML} found that for the SET HASDM density database, the use of NLPD as opposed to MSE did not result in significantly different prediction errors.

\subsection*{Batch Size for Tuners and Further Training}
\setcounter{table}{0}
\renewcommand{\thetable}{C\arabic{table}}

Batch size is the number of samples used to average the loss during fitting. With too small of a batch size, the model can update its weights often to specific batches of the data making it unstable. With too large of a batch size, the model can have trouble learning due to the over-generalization of the losses. Choosing a batch size for a given application requires trial and error as there is no universal choice. Table \ref{t:batch_size} shows the batch sizes used for the tuners and subsequent training for each dataset.

\begin{table}[htb!]
	\fontsize{10}{10}\selectfont
    \caption{Batch sizes used to tune and train the three ML models in this work.}
   \label{t:batch_size}
        \centering 
   \begin{tabular}{ c | c | c } % Column formatting, 
      \hline 
           & \textbf{JB2008-ML / HASDM-ML} & \textbf{\;\;\;\;\;CHAMP-ML\;\;\;\;\;} \\ \hline 
           \textbf{Tuner} & 2\textsuperscript{8} = 256 & 2\textsuperscript{12} = 4,096 \\ \hline
           \textbf{Further Training} & N/A & 2\textsuperscript{17} = 131,072 \\ \hline
   \end{tabular}
\end{table}

Batch sizes in Table \ref{t:batch_size} were chosen from trial and error. The JB2008-ML and HASDM-ML batch size is the smallest partially due to the size of the dataset (35,064 training samples). For the CHAMP-ML tuner, we use 1 million samples and therefore have a larger batch size. The batch size for training the final CHAMP-ML model using the best architecture from the tuner is significantly larger, because we have over 20 million training samples and found this leads to the most stable training process.
\end{document}